\documentclass[12pt]{article}
\usepackage{amsmath}
\usepackage{graphicx,psfrag,epsf}
\usepackage{enumerate}
\usepackage{natbib}
\usepackage{url} 

\usepackage{natbib,bm}
\usepackage{amsmath,amssymb}
\usepackage{graphicx}
\usepackage{epsfig}
\usepackage{epstopdf}
\usepackage{rotating}
\usepackage{multirow}
\usepackage{multicol}
\usepackage[small]{caption}
\usepackage{subfig}
\usepackage{color}
\usepackage{mathtools}
\usepackage{enumerate}
\usepackage{pdflscape}
\usepackage{longtable}
\usepackage{pifont}

\usepackage{setspace}

\newcommand{\nmathbf}{\bm}

\def\bfh{\nmathbf h}

\def\bfs{\nmathbf s}

\def\bfx{\nmathbf x}

\def\bfbeta   {\nmathbf \beta}

\def\boldfacefake#1{\kern-4pt
   \hbox{ \mathsurround=0pt
   \hbox to 0.4pt{$#1$\hss}\hbox to 0.4pt{$#1$\hss}\hbox {$#1$}}}

\newcommand{\btable}{\begin{table}[h]\centering}
\newcommand{\etable}{\end{table}}
\newcommand{\bt}{\begin{parag}\small \let\b=\nsb \let\sb=\nssb \begin{tabular}}
\newcommand{\et}{\end{tabular}\let\b=\nb \let\sb=\nsb\end{parag}}

\newenvironment{parag}{\par}{\par}

\newcommand{\be}{\begin{eqnarray}}
\newcommand{\ee}{\end{eqnarray}}
\newcommand{\ba}{\begin{eqnarray*}}
\newcommand{\ea}{\end{eqnarray*}}

\newcommand{\reals}{\mbox{\rm I\kern-.20em R}}
\newcommand{\sreals}{\mbox{\small \rm I\kern-.20em R}}

\newcommand{\blind}{0}

\begin{document}

\def\spacingset#1{\renewcommand{\baselinestretch}%
{#1}\small\normalsize} \spacingset{1}


\if0\blind
{
  \title{\bf Bayesian Survival Analysis Using Gamma Processes with Adaptive Time Partition}
  \author{Yi Li\\
    {\small Department of Mathematics, Northeastern University}\\
\\
    Sumi Seo \\
    {\small Department of Mathematics, Northeastern University}\\
\\
    Kyu Ha Lee \\
    {\small Department of Nutrition, Department of Epidemiology, Department of Biostatistics,}\\
    {\small Harvard T.H. Chan School of Public Health}}
  \maketitle
} \fi

\if1\blind
{
  \bigskip
  \bigskip
  \bigskip
  \begin{center}
    {\LARGE\bf Title}
\end{center}
  \medskip
} \fi

\smallskip
\begin{abstract}
In Bayesian semi-parametric analyses of time-to-event data, non-parametric process priors are adopted for the baseline hazard function or the cumulative baseline hazard function for a given finite partition of the time axis. However, it would be controversial to suggest a general guideline to construct an optimal time partition. While a great deal of research has been done to relax the assumption of the fixed split times for other non-parametric processes, to our knowledge, no methods have been developed for a gamma process prior, which is one of the most widely used in Bayesian survival analysis. In this paper, we propose a new Bayesian framework for proportional hazards models where the cumulative baseline hazard function is modeled \emph{a priori} by a gamma process. A key feature of the proposed framework is that the number and position of interval cutpoints are treated as random and estimated based on their posterior distributions. 
\end{abstract}

\vfill

\newpage
\spacingset{1.45} 
\section{Introduction} \label{sec:intro}

Non-parametric and semi-parametric Bayesian analysis of time-to-event data have been well accepted in practice because they enable more flexible modeling strategy with fewer assumptions. While semi-parametric Cox proportional hazards models \citep{cox1972regression} avoid modeling the baseline hazard function by maximizing the partial likelihood, the Bayesian paradigm requires explicit parametrization of the baseline hazard function. To this end, various types of non-parametric prior processes have been developed, including gamma process \citep{kalbfleisch1978non}, beta process \citep{hjort1990nonparametric}, and Dirichlet process \citep{gelfand1995bayesian}, which can be used to model the cumulative baseline hazard or distribution functions. A broad overview can be found in \cite{ibrahim2005bayesian}.

Gamma process is one of the most widely used non-parametric process priors to model the cumulative baseline hazard in Bayesian proportional hazards models \citep{kalbfleisch1978non, clayton1991monte, sinha1999bayesian, sinha2003bayesian, sinha2015bayesian}. It has been adopted in many applications including multivariate methods \citep{dunson2004selecting, sen2010bayesian, cai2010bayesian, sreedevi2012semiparametric, ouyang2013bayesian} and variable selection methods \citep{lee2011bayesian, savitsky2011variable, gu2013bayesian, zhao2014bayesian, lee2015survival}, and other survival models  \citep{gelfand2003bayesian, cho2009bayesian, zhao2015dirichlet}. In the absence of the prior information, in the survival models where Gamma process prior is used for cumulative baseline hazard functions \citep{burridge1981empirical, lee2011bayesian, savitsky2011variable, gu2013bayesian, zhao2014bayesian, lee2015survival}, the time partition is specified either based on uniquely ordered failure times or equi-length intervals conditional on the prespecified number of interval cutpoints. The use of equally spaced intervals may lead to the time partition where some of the intervals have few failures. The possibility increases as the number of splits is set to a large value. On the other hand, setting the partition such that each interval contains the same number of failures may result in oversimplified estimation of baseline hazard for a long period time. Thus, fixing the partition of the baseline hazard timescale may impose unreasonable structure on the model and consequently standard error estimates may not reflect additional uncertainty associated with the number and position of the split times \citep{Haneuse2008}. Therefore, it is very important to specify the optimal time partition in the Bayesian semiparametric analysis of time-to-event data.

While there have been studies on methods to relax the assumption of fixed time partition for models with other non-parametric processes \citep{arjas1994nonparametric, Haneuse2008, Lee2014bscr}, to our knowledge, no such methods have been developed for the gamma process prior for cumulative baseline hazard function. In this paper, we propose a Bayesian framework that allows the number and position of interval splits to be determined by the data. We present an efficient computational scheme to fit the proposed models based on a reversible jump Markov chain Monte Carlo (MCMC) algorithm \citep{Green95reversiblejump}. 

The remainder of the paper is organized as follows. Section \ref{sec:model} describes the proposed Bayesian analysis framework including specification of prior distributions. Section \ref{sec:computation} provides a detailed computational algorithm for obtaining samples from the joint posterior and two metrics for comparing goodness of fit across model specifications. Section \ref{sec:simulation} presents simulation studies under the four different scenarios to assess performance of our proposed methods. In Section \ref{sec:application}, we apply the proposed framework to four different real-life survival data. Section \ref{sec:discussion} concludes with discussion.

\section{A Bayesian Framework for Proportional Hazards Models} \label{sec:model}

\subsection{Model Specification and Observed Data Likelihood}
Let $T$ denote the survival times of individuals in a population. Under the Cox proportional hazards model \citep{cox1972regression}, the hazard at time $t$ for an individual whose covariate vector is $\bfx$ is given by
\be
	h(t| \bfx) = h_0(t)\exp(\bfx^{\top}\bfbeta),\label{eqn:PH}
\ee
where $h_0(t)$ is the baseline hazard function and $\bfbeta$ is a vector of regression coefficients. In the Bayesian paradigm, one is required to specify the baseline hazard function. We consider a non-parametric specification by taking the baseline hazard function to be a flexible mixture of piecewise constant functions \citep{mckeague2000bayesian}. To this end, we construct a finite partition of the time axis, $\bfs = \{s_0<s_1<s_2<...<s_{J+1}\}$, with $ s_0\equiv 0$, $s_{\textrm{max}}\equiv s_{J+1} > t_i$ for all $i=1,...,n$. The survival time $t_i$ of the $i^{th}$ subject falls in one of the $J+1$ disjoint intervals. Given the partition ($J$, $\bfs$), we assume
\be
	h_0(t) =  \sum_{j=1}^{J+1} \frac{h_j }{s_j-s_{j-1}}1_{[s_{j-1}<t\leq s_{j}]}, \label{eqn:cumParam}
\ee
where $h_j$ is the increment in the cumulative baseline hazard $H_0(t)$ in the $j^{\textrm{th}}$ interval, that is $h_j=H_0(s_j)-H_0(s_{j-1})$. This specification is referred to as a piecewise exponential model (PEM) \citep{ibrahim2005bayesian}. Let $C_i$ denote the right censoring time for the $i^{\textrm{th}}$ subject and let $Y_i = \min(T_i, C_i)$, $\delta_i=1$ if $Y_i=T_i$ and $0$ otherwise. Furthermore, let  $D_i = \{y_i, \delta_i, \bfx_i\}$ denote the observed data for the $i^{\textrm{th}}$ subject. For a given specification of (\ref{eqn:PH}) and (\ref{eqn:cumParam}), the observed data likelihood as a function of the unknown parameters $\Phi=(\bfbeta, \bfh=\{h_j\}_{j=1}^{J+1})$, is given by
\be
	&&L(D|\Phi)= \prod_{i=1}^n L(D_i|\Phi)\nonumber \\
	&=&\prod_{i=1}^{n} \prod_{j=1}^{J+1} \left[\left\{\frac{h_j}{s_j-s_{j-1}}\exp(\bfx_i^{\top}\bfbeta)\right\}^{\delta_i } \exp\left\{-\sum_{g=1}^j \frac{h_g\Delta_j(y_i)}{s_g-s_{g-1}}\right\}\right]^{1_{[s_{j-1}<y_i\leq s_j]}}                            \nonumber \\
	&=&\prod_{j=1}^{J+1}\left[\left(\frac{h_j}{s_j-s_{j-1}}\right)^{d_j}\left\{\prod_{k\in D_j}\exp(\bfx_k^{\top}\bfbeta)\right\} \exp\left\{-\sum_{l\in R_j} \Delta_j(y_l)\frac{h_j}{s_j-s_{j-1}}\exp(\bfx_l^{\top}\bfbeta) \right\}\right], \nonumber
\ee
where $\mathcal{R}_j$ is the set of subjects at risk, $\mathcal{D}_j$ is the set of subjects having failures in the $j^{th}$ interval, $d_j=|\mathcal{D}_j|$, and $\Delta_j(y)=\max\left\{0, \min(y, s_j) - s_{j-1}\right\}$.

\subsection{Hyperparameters and Prior Distributions}

In the proposed Bayesian framework, we use a gamma process prior for the cumulative baseline hazard function $H_0(t)$ \citep{kalbfleisch1978non}:
\be
    H_0\sim \mathcal{G}\mathcal{P} \left(c_0H^{\ast}, c_0\right), \label{eqn:GP}
\ee 
where $H^{\ast}(t)$ is an increasing function with $H^{\ast}(0)=0$, and $c_0$ is a positive constant. The specified $H^{\ast}$ can be viewed as an initial guess of $H_0$ and $c_0$ is a specification of weight or confidence attached to that guess. In general, $H^{\ast}$ is taken to be a known parametric function, such as a Weibull distribution. Specifically, we consider $H^{\ast}(t)=\eta_0t^{\kappa_0}$, where $(\eta_0, \kappa_0)$ are hyperparameters. The gamma process prior (\ref{eqn:GP}) implies that $h_j$'s follow independent gamma distributions:
\be
	    h_j\sim \textrm{Gamma}(c_0\eta_0(s_{j})^{\kappa_0}-c_0\eta_0(s_{j-1})^{\kappa_0}, ~c_0). \nonumber
\ee
Thus, the hyperparameters $(\eta_0, \kappa_0, c_0)$ for $h_j$ consist of a specified parametric cumulative hazard function $H^*(t)=\eta_0t^{\kappa_0}$ evaluated at the splits of the time partition, and a positive scalar $c_0$ quantifying the degree of prior confidence in $H^*(t)$. 

We note that it is often challenging to determine the number of split times $J$ and their positions $\bfs$. Thus the choice is often based on heuristic considerations. We avoid reliance on a fixed partition of the time axis by permitting the partition ($J$, $\bfs$) to vary and to be updated via a reversible jump MCMC scheme \citep{Green95reversiblejump, mckeague2000bayesian,Haneuse2008}. Specifically, $J$ is drawn from a Poisson distribution with rate $\alpha$ but restricted to be $J\leq J_{\max}$. Conditional on the number of splits, we take the positions $\bfs$ to be \emph{a priori} distributed as the even-numbered order statistics. This can help decrease the probability of creating a new interval containing no events, which could make estimation of the baseline hazard in the interval difficult. Finally, we adopt a noninformative flat prior on the real line for regression parameters $\bfbeta$. To summarize, our prior specification is as follows.
\be
h_j & \sim&\textrm{Gamma}\left(c_0\eta_0(s_{j})^{\kappa_0}-c_0\eta_0(s_{j-1})^{\kappa_0}, ~c_0\right),  ~j=1,\cdots,J+1 \nonumber \\
J&\sim& \textrm{Poisson}(\alpha) \nonumber\\
\pi(\bfs|J) & \propto& \frac{(2J+1)!}{(s_{J+1})^{2J+1}}\prod_{j=1}^{J+1}(s_j-s_{j-1}) \nonumber \\
\pi(\bfbeta) & \propto& 1 \nonumber
\ee

\section{Posterior Inference and Model Comparison} \label{sec:computation}

\subsection{Markov Chain Monte Carlo}

To perform posterior estimation and inference, we use a random-scan Gibbs sampling algorithm to generate samples from the joint posterior distribution. Since updating the time partition ($J$, $\bfs$) requires a change in the dimension of the parameter space, we develop our computational scheme based on a Metropolis-Hastings-Green (MHG) step \citep{Green95reversiblejump}. A detailed description of the complete algorithm, together with all necessary full conditional posterior distributions, is provided. 

In the first step of the random-scan Gibbs sampling scheme, we select a move from the following four possible choices:\\

\noindent move {\bf RP}: updating regression parameters $\bfbeta$\\
move {\bf BH}: updating parameters for the cumulative baseline hazard $\bfh$\\
move {\bf BI}: creating a new split time\\
move {\bf DI}: removing a split time\\

\noindent If $J$ is the number of time splits at the current iteration, the probabilities for move ${\bf BI}$ and ${\bf DI}$ are
\begin{align}
  \pi_{BI}^{J} & =\rho\min\left\{1,\frac{\textrm{Poisson}(J+1|\alpha)}{\textrm{Poisson}(J|\alpha)}\right\} =\rho\min\left\{1, \frac{\alpha}{J+1}\right\} \nonumber \\
  \pi_{DI}^{J} & =\rho\min\left\{1,\frac{\textrm{Poisson}(J-1|\alpha)}{\textrm{Poisson}(J|\alpha)}\right\} =\rho\min\left\{1, \frac{J}{\alpha}\right\}. \nonumber
\end{align}
$\rho$ is determined so that $\pi_{BI}^{J}+\pi_{DI}^{J}<C$ and $C<1$ for $J=1,\ldots,J_{\max}$. $J_{\max}$ is the preassigned upper limit on the number of time splits and we set $\pi_{BI}^{J_{\max}}=0$. For the remaining moves,
\be
	\pi_{RP}=\pi_{BH}=\frac{1-\pi_{BI}^{J}-\pi_{DI}^{J}}{2} \nonumber
\ee

\noindent\underline{\bf Move RP}

The full conditional posterior distribution for $\beta_m$, $m=1,\cdots,p$, is given by

\be\label{betajfullconditional}
  \pi(\beta_m|\bfbeta^{(-m)},\bfh)\propto \prod_{j=1}^{J+1}\left\{\prod_{k\in D_j}\exp(x_{k,m}\beta_m)\right\} \exp\left\{-\sum_{l\in R_j} \Delta_j(y_l)\frac{h_j}{s_j-s_{j-1}}\exp(x_{l,m}\beta_m) \right\}, \nonumber
\ee
where $\bfbeta^{(-m)}$ is $\bfbeta$ without the element $\beta_m$. Since the full conditional does not have standard form, we update each of regression coefficient using a random walk Metropolis-Hastings (MH) algorithm.\\

\noindent\underline{\bf Move BH}

The full conditional posterior distribution for $h_j$, $j=1,\cdots,J+1$, is given by
\be\label{hjfullconditional}
  \pi(h_j|\bfh^{(-j)},\bfbeta)\propto h_j^{d_j + c_0\eta_0(s_{j})^{\kappa_0}-c_0\eta_0(s_{j-1})^{\kappa_0}-1} \exp\left\{-h_j\left(\sum_{l\in R_j} \frac{\Delta_j(y_l)e^{\bfx_l^{\top}\bfbeta}}{s_j-s_{j-1}} +c_0\right)\right\}. \nonumber
\ee
Since the full conditional does not have standard form, we update $h_j$'s using a random walk MH algorithm.\\

\noindent\underline{\bf Move BI}

For a given partition $(J,\bfs)$, the cumulative baseline function $H_0(t$)=$\sum_{j=1}^{J+1}\frac{\Delta_j(t)}{s_j-s_{j-1}}h_j$. Updating this specification requires generating a proposal split and then deciding whether or not to accept the proposal. First, we proceed by selecting a proposal split time $s^*$ uniformly from among the observed event times that are not included in the current partition. Suppose $s^*$ falls in the interval ($s_{j-1}$, $s_{j}$] in the current partition. That is, the induced proposal partition can be written as
\be
   (0=s_0,\cdots,s_{j-1},s^*,s_{j},\cdots,s_{J+1}=s_{\max}) \equiv  (0=s_0^*,\cdots,s_{j-1}^*,s_j^*,s_{j+1}^*,\cdots,s_{J+2}^*=s_{\max}) \nonumber
\ee
Second, we calculate $h_{j}^*$ and $h_{j+1}^*$ of the two new intervals created by the split at time $s^*$. To ensure the old height, $\frac{h_j}{s_j-s_{j-1}}$, is a compromise of the two new heights, $\frac{h_j^*}{s^*-s_{j-1}^*}$ and $\frac{h_{j+1}^*}{s_{j+1}^*-s^*}$, the former is taken to be the weighted mean of the latter:
\be
	(s^*-s_{j-1}^*)\log\frac{h_{j}^*}{s^*-s_{j-1}^*} + (s_{j+1}^*-s^*)\log\frac{h_{j+1}^*}{s_{j+1}^*-s^*} = (s_j-s_{j-1})\log\frac{h_{j}}{s_j-s_{j-1}}. \nonumber
\ee
We define the multiplicative perturbation $\frac{h_{j}^*(s_{j+1}^*-s^*)}{h_{j+1}^*(s^*-s_{j-1}^*)}=\frac{1-U}{U}$, where $U\sim$Uniform(0, 1). Then the new $h_{j}^*$ and $h_{j+1}^*$ are given by
\be
	h_j^* &=&  h_j\frac{s^*-s_{j-1}}{s_j-s_{j-1}}\left(\frac{1-U}{U}\right)^{\frac{s^*-s_{j-1}}{s_j-s_{j-1}}} \nonumber \\
	h_{j+1}^* &=&  h_j\frac{s_j-s^*}{s_j-s_{j-1}}\left(\frac{U}{1-U}\right)^{\frac{s_j-s^*}{s_j-s_{j-1}}}. \nonumber
\ee
The Jacobian of the above system is 
\be
\left |
  \begin{array}{cc}
    \frac{\partial h_j^*}{\partial h_j} & \frac{\partial h_j^*}{\partial U} \\
    \frac{\partial h_{j+1}^*}{\partial h_j} & \frac{\partial h_{j+1}^*}{\partial U} \\
  \end{array}
\right |=h_j\left[\frac{(s^*-s_{j-1})(s_j-s^*)}{\left\{(1-U)^{(s^*-s_{j-1})}U^{(s_j-s^*)}\right\}^{\frac{2}{s_j-s_{j-1}}}}\right]. \nonumber
\ee
Finally, the acceptance probability in the MHG step is computed as the product of the likelihood ratio, prior ratio, proposal ratio, and Jacobian:

\begin{enumerate}[i)]
	\item Likelihood ratio: $L(\bfbeta,\bfh^*)/L(\bfbeta,\bfh)$
	\item Prior ratio:
	\be
		&&\frac{\textrm{Poisson}(J+1|\alpha)}{\textrm{Poisson}(J|\alpha)}\times \frac{\prod_{g=j}^{j+1}\textrm{Gamma}(h_g^* | c_0\eta_0(s_{g}^*)^{\kappa_0}-c_0\eta_0(s_{g-1}^*)^{\kappa_0}, ~c_0)}{\textrm{Gamma}(h_j | c_0\eta_0(s_{j})^{\kappa_0}-c_0\eta_0(s_{j-1})^{\kappa_0}, ~c_0)} \nonumber \\
		&\times&\frac{(2J+3)(2J+2)(s^*-s_{j-1})(s_j-s^*)}{s_{\max}^2(s_j-s_{j-1})} \nonumber
	\ee 
	\item Proposal ratio:
	 \be
		\frac{\pi_{DI}\sharp\{y_i:\delta_i=1\}}{\pi_{BI}(J+1)\textrm{Uniform}(U|0, 1)} \nonumber
	\ee
	\item Jacobian:
	\be
		h_j\left[\frac{(s^*-s_{j-1})(s_j-s^*)}{\left\{(1-U)^{(s^*-s_{j-1})}U^{(s_j-s^*)}\right\}^{\frac{2}{s_j-s_{j-1}}}}\right] \nonumber
	\ee
\end{enumerate}

\noindent\underline{\bf Move DI}

The acceptance probability for the corresponding reverse move has the same form with the appropriate change of labelling of the partitions and variables, and the ratio terms inverted. Suppose we remove a randomly selected split time $s_j$. The proposal partition of time axis consists of $J$ time splits as follows:
\be
   (0=s_0,\cdots,s_{j-1},s_{j+1},\cdots,s_{J+1}=s_{\max}) \equiv(0=s_0^*,\cdots,s_{j-1}^*,s_j^*,s_{j+1}^*,\cdots,s_{J}^*=s_{\max}) \nonumber
\ee
As similarly done for the move BI,
\be \label{eq:2}
(s_j-s_{j-1})\log\frac{h_{j}}{s_j-s_{j-1}} + (s_{j+1}-s_j)\log\frac{h_{j+1}}{s_{j+1}-s_j} &=& (s_j^*-s_{j-1}^*)\log\frac{h_{j}^*}{s_j^*-s_{j-1}^*}  \nonumber\\
 \frac{h_{j}(s_{j+1}-s_j)}{h_{j+1}(s_j-s_{j-1})} &=& \frac{1-U^*}{U^*}\nonumber
\ee
Therefore, the four components of the acceptance probability can be written as follows:
\begin{enumerate}[i)]
	\item Likelihood ratio: $L(\bfbeta,\bfh^*)/L(\bfbeta,\bfh)$
	\item Prior ratio:
	\be
		&&\frac{\textrm{Poisson}(J-1|\alpha)}{\textrm{Poisson}(J|\alpha)}\times \frac{\textrm{Gamma}(h_j^* | c_0\eta_0(s_{j}^*)^{\kappa_0}-c_0\eta_0(s_{j-1}^*)^{\kappa_0}, ~c_0)}{\prod_{g=j}^{j+1}\textrm{Gamma}(h_g | c_0\eta_0(s_{g})^{\kappa_0}-c_0\eta_0(s_{g-1})^{\kappa_0}, ~c_0)} \nonumber \\
		&\times&\frac{s_{\max}^2(s_{j+1}-s_{j-1})}{(2J+1)2J(s_j-s_{j-1})(s_{j+1}-s_j)} \nonumber
	\ee 
	\item Proposal ratio:
	\be
		\frac{\pi_{BI}J}{\pi_{DI}\sharp\{y_i:\delta_i=1\}} \nonumber
	\ee	
	\item Jacobian:
	\be
		\frac{1}{h_j^*}\left[\frac{\left\{(1-U^*)^{(s_j-s_{j-1})}(U^*)^{(s_{j+1}-s_j)}\right\}^{\frac{2}{s_{j+1}-s_{j-1}}}}{(s_j-s_{j-1})(s_{j+1}-s_j)}\right] \nonumber
	\ee	
\end{enumerate}

\subsection{Model Comparison Criteria}

In practice, analysts must balance model complexity with limitations of available information in the data. To this end, we develop two model comparison metrics based on the deviance information criterion (DIC) \citep{spiegelhalter2002bayesian} and the log-pseudo marginal likelihood (LPML) statistics \citep{geisser1979predictive}. Various DIC constructions and extensions have been compared and tested in the case of mixtures of distributions and random effect models \citep{celeux2006deviance}. Since our proposed framework involves the PEM structure, we consider the DIC$_3$ measure as suggested in \cite{celeux2006deviance} and estimate the quantity by using the following Monte Carlo approximation:
\be
	\widehat{\mbox{DIC}}_{\mbox{\tiny GP}}\ =\ -\frac{4}{R}\sum_{r=1}^R\log\left\{\prod_{i=1}^n L (D_i | \Phi^{(r)})\right\}\ +\ 2\log\left\{ \prod_{i=1}^n\frac{1}{R}\sum_{r=1}^R L (D_i | \Phi^{(r)})\right\}.\nonumber
\ee
\noindent $\Phi^{(r)}$ denotes the value of $\Phi$ at the $r^{\textrm{th}}$ MCMC iteration, $r$=$1,\ldots,R$. Note, a model with smaller DIC$_{\mbox{\tiny GP}}$ indicates a better fit of the model for the data. A general rule of thumb for model comparison is to consider differences of less than 2 to be negligible, differences between 2 and 6 to be indicative of positive support for the model with the lower value and differences greater than 6 to be strong support in value of the model with the lower value \citep{spiegelhalter2002bayesian}. 

The LPML criterion is computed as $\sum_{i=1}^n\log \textrm{CPO}_i$, where the subject-specific conditional predictive ordinates (CPO) is given by $\textrm{CPO}_i=L (D_i | D^{-(i)})$. Following \cite{chen2012monte}, CPO can be approximated via a Monte Carlo estimator:
\be
	\widehat{\textrm{CPO}_i}\  =\left\{\frac{1}{R}\sum_{r=1}^R L (D_i | \Phi^{(r)})^{-1}\right\}^{-1}. \label{CPO} \nonumber
\ee
Since CPO$_i$ is the marginal posterior predictive density of the outcome for the $i^{th}$ subject given a dataset excluding the subject, a larger value of LPML indicates a better fit to the data. For practical interpretation, one can compute the so-called pseudo-Bayes factor (PBF) for two models by exponentiating difference in their LPML values \citep{hanson2006inference}.

\begin{table}[ht]
\centering
\caption{A summary of four simulation scenarios explored in Section \ref{sec:simulation}.\label{tab:simSetting}}
\begin{tabular}{c c c c}
\hline
Scenario		& $n$	&	Censoring &Distribution of the			\\
 			&		& rate & baseline hazard function 	\\
\hline
1 & 300 & 30\% & Weibull 	\\
2 & 300 & 30\% & Piecewise linear \\
3 & 300 & 50\% & Weibull 	\\
4 & 100 & 30\% & Weibull 	\\
\hline
\end{tabular}
\end{table}

\begin{figure}[ht]
\centering
\includegraphics[width = 2.2in]{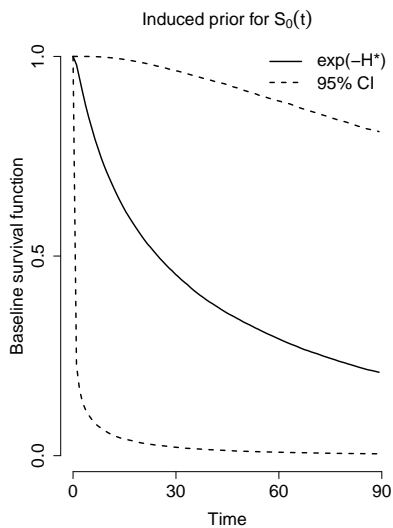}
\caption{Induced baseline survival function $S_0(t)$ based on the hyperparameters considered in Section \ref{sec:simulation} and \ref{sec:application}.}\label{figure:priorSurv}
\end{figure}

\section{Simulation Studies}\label{sec:simulation}

We refer to the proposed Bayesian framework as a PEM-gamma process model with a reversible jump MCMC scheme (GP-RJ). The overarching goals of the simulation studies are to investigate the small sample operating characteristics of the proposed GP-RJ framework and compare the performance with a PEM-gamma process model with a fixed time partition with equally spaced intervals (GP-EQ). GP-EQ model can be implemented by running GP-RJ with $\pi_{BI}$=$\pi_{DI}$ set to $0$. 


\subsection{Setting}

We consider four data scenarios that vary in terms of sample size, censoring rates, and the true underlying baseline hazard distribution. The simulation setting is similar to that in \citep{lee2016hierarchical} and a summary of the scenarios is provided in Table \ref{tab:simSetting}. In Scenarios 1, 3, and 4, the baseline hazard function is set to correspond to the hazard of a Weibull($0.8$, $0.05$). To evaluate the performance of the model when the true baseline hazard function is not monotone increasing or decreasing function like a Weibull, we consider a piecewise linear hazard function in Scenario 2: $h_0(t)=\{-(k-b)t/40 + b\}1_{[t\leq 40]}$ + $\{(3k-b)/2+(k-b)t/80\}1_{[t> 40]}$ with $b=0.1, k=0.0005$. In order to carry out investigation on the effect of sample size and availability of information in the data, we consider a higher censoring rate ($50\%$) in Scenario 3 and a smaller sample size ($n=100$) in Scenario 4.

For each of the four scenarios we generated 500 simulated datasets under the model outlined in Table \ref{tab:simSetting}. We specified that the hazard function depended on three covariates: $x_{i,1}$, $x_{i,2}\sim$ Normal(0, 1), and $x_{i,3}\sim$ Bernoulli(0.5). The regression coefficients were set to $(\beta_1, \beta_2, \beta_3)=(0.5, 0.8, -0.5)$. We used the \texttt{simSurv} function in the \texttt{SemiCompRisks} \texttt{R} package \citep{alvares2019semicomprisks} to simulate the survival data.

\subsection{Analyses and Specification of Hyperparameters}

For each of the 500 datasets generated under each of the four scenarios we fit GP-RJ and GP-EQ models. For GP-RJ model, we set the prior Poisson rate on the number of split times to be $\alpha=10$. We specified that the time partition for GP-EQ model was constructed with 11 equi-length intervals ($J=10$). For both models, we set $(\eta_0, \kappa_0, c_0) = (0.2, 0.5, 1)$. The induced baseline survival function $S_0(t)=\exp(-H_0(t))$ based on this choice of hyperparameters was given in Figure \ref{figure:priorSurv}. We ran two independent chains for a total of 1 million scans each with the first half taken as burn-in. We computed the Gelman-Rubin potential scale reduction factors (PSRF) \citep{gelman2013bayesian} to assess convergence, specifically requiring the statistics to be less than 1.05 for all model parameters.

\begin{table}[ht]
\centering
\caption{Estimated percent bias (PB), coverage probability (CP), and average relative width of 95\% credible intervals (RW) for $\bfbeta$ for two analyses described in Section \ref{sec:simulation}, across four simulation scenarios given in Table \ref{tab:simSetting}. For RW, the GP-RJ was taken as the referent and throughout values are based on results from 500 simulated datasets. \label{tab:simReg}}
\begin{tabular}{ccr c cc c cc c cc}
  \hline
	 		&&   			&& \multicolumn{2}{c}{PB} && \multicolumn{2}{c}{CP} && \multicolumn{2}{c}{RW}  \\ \cline{5-6}\cline{8-9}\cline{11-12}   
Scenario	 	&& True	&& GP-RJ   & GP-EQ 		&& GP-RJ & GP-EQ 				&& GP-RJ & GP-EQ 		\\
  \hline
 		& $\beta_{1}$ & 0.5 	&& 1.8 & 2.5 && 0.95 & 0.94 && 1.00 & 1.00 \\ 
1 		& $\beta_{2}$ & 0.8 	&& 2.8 & 3.6 && 0.94 & 0.93 && 1.00 & 1.00 \\ 
 		& $\beta_{3}$ & -0.5	&& 2.2 & 2.7 && 0.95 & 0.96 && 1.00 & 1.00 \\ 
  \hline
 		& $\beta_{1}$ & 0.5 	&& 1.7 & 2.0 && 0.94 & 0.95 && 1.00 & 1.00 \\ 
2 		& $\beta_{2}$ & 0.8 	&& 2.5 & 2.8 && 0.94 & 0.93 && 1.00 & 0.99 \\ 
 		& $\beta_{3}$ & -0.5	&& -0.4 & -0.6 && 0.95 & 0.96 && 1.00 & 1.00 \\ 
    \hline
 		& $\beta_{1}$ & 0.5 	&& 1.7 & 2.5 && 0.94 & 0.93 && 1.00 & 1.00 \\ 
3 		& $\beta_{2}$ & 0.8 	&& 3.0 & 3.8 && 0.94 & 0.94 && 1.00 & 1.00 \\ 
 		& $\beta_{3}$ & -0.5	&& 2.6 & 3.3 && 0.95 & 0.95 && 1.00 & 1.00 \\ 
    \hline
 		& $\beta_{1}$ & 0.5 	&& 6.2 & 6.2 && 0.95 & 0.94 && 1.00 & 1.00 \\ 
4 		& $\beta_{2}$ & 0.8 	&& 3.3 & 3.3 && 0.94 & 0.94 && 1.00 & 1.00 \\ 
 		& $\beta_{3}$ & -0.5	&& 0.9 & 0.8 && 0.95 & 0.95 && 1.00 & 1.00 \\ 		
   \hline
\end{tabular}
\end{table}

\begin{figure}[hp]
\centering
\includegraphics[width = 5.2in]{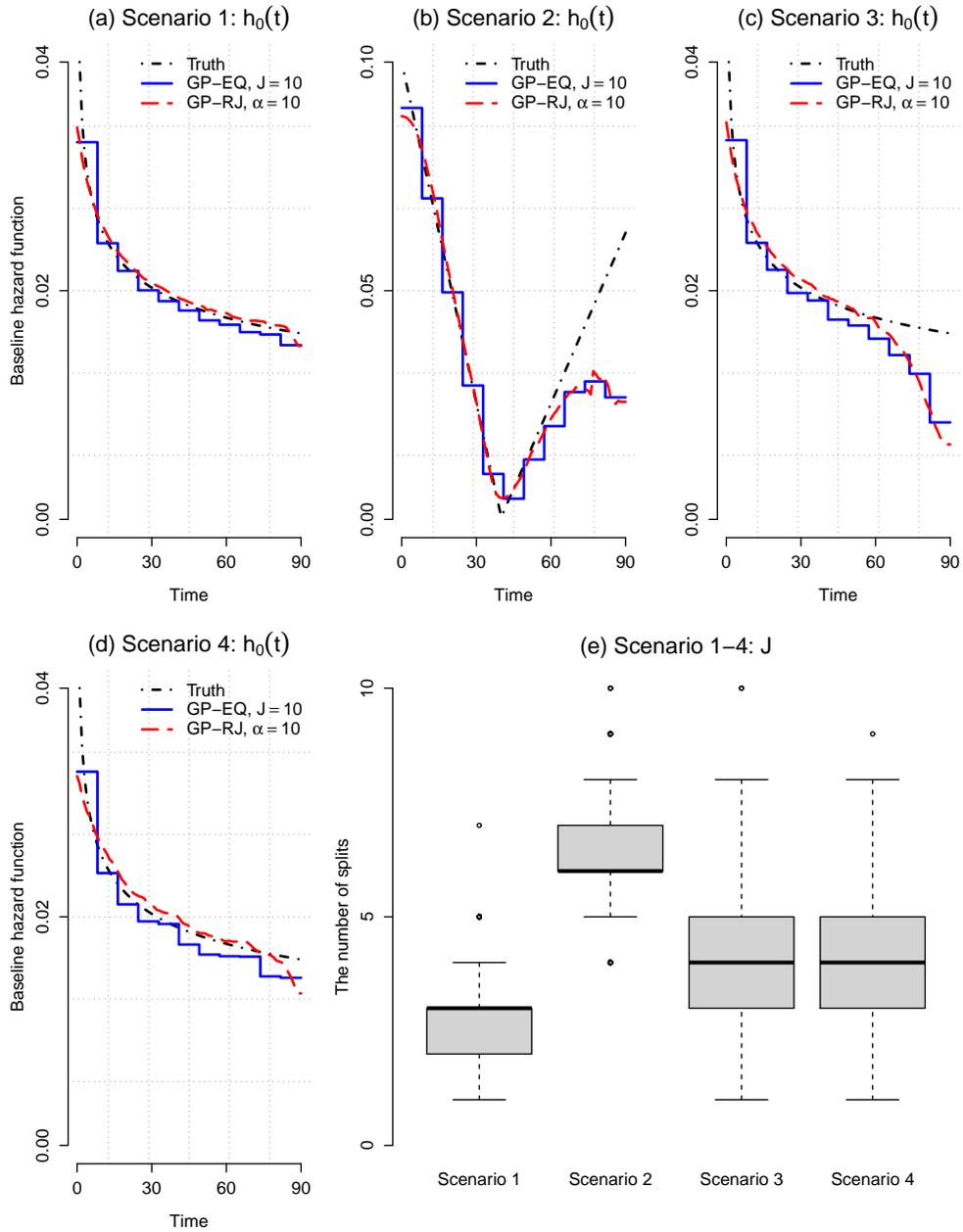}
\caption{Simulation studies: (a)-(d) Estimated baseline hazard function, $h_0(t)$, for two analyses described in Section \ref{sec:simulation}; (e) The posterior distribution of $J$ from the analysis under the GP-RJ model.} \label{figure:simBH}
\end{figure}

\subsection{Results}

In Figure \ref{figure:simBH} (a)-(d), we presented the mean estimated baseline hazard function under the four scenarios using GP-RJ and GP-EQ models. Both models performed well in terms of estimation of the baseline hazard function. In Scenarios 2 and 3, the baseline hazard from the analyses tended to be underestimated for $t>60$. This was due to relatively fewer events being generated for the later time period under the two simulation scenarios. Note that the proposed GP-RJ model successfully smoothed out the process while the GP-EQ resulted in a piecewise constant baseline hazard as the conventional PEM model. In Figure \ref{figure:simBH} (e), we presented the distribution of the estimated number of time splits ($J$) for GP-RJ. While providing a smooth estimated baseline hazard function function like the truth, GP-RJ with $\alpha=10$ utilized a smaller number of split times comparing to GP-EQ with $J=10$ across the four scenarios. In addition, it was shown that the more complicated the underlying hazard function was (Scenario 2) and the less the information was available from the data (Scenarios 3 and 4), the more split times GP-RJ tended to create.

Table \ref{tab:simReg} indicated that the proposed GP-RJ also performed well in terms of estimation and inference for $\bfbeta$. It was shown that percent bias was small across the board, and the estimated coverage probabilities were all close to the nominal 0.95. While GP-EQ yielded point estimates of $\bfbeta$ that were slightly more biased comparing to GP-RJ, the difference between two models was not significant because of the nature of the proportional hazards assumption.

\section{Applications} \label{sec:application}

In this section, we demonstrate the practical utility of our proposed models using four different time-to-event datasets. The four datasets are Dutch male laryngeal cancer data (DLC) \citep{kardaun1983statistical}, German breast cancer data (GBC) \citep{schumacher1994randomized}, Netherlands Cancer Institute breast cancer data (NBC) \citep{van2002gene}, and American Association for Cancer Research breast cancer data (ABC) \citep{shapiro1993comparison}. We provide a summary of the four datasets in Table \ref{tbl:dataSummary}.

\begin{table}[ht]  
 \begin{center}
  \caption{ A summary of the datasets analyzed in Section \ref{sec:application} \label{tbl:dataSummary} }
\begin{tabular}{l l c  c  c}
\hline
  Data ID           & Description 	& $n$           	& $p$ 	& Censoring\\
			&  			&            		&  		& rate\\
\hline
    Data 1. DLC   &	Dutch male laryngeal cancer data	& 90&    3   & 44\%   \\
    Data 2. GBC   &   German breast cancer data       & 686 &    8   & 56\%   \\
    Data 3. NBC   &   NCI breast cancer data          & 144 &   5      &  67\%\\
    Data 4. ABC   &    AACR breast cancer data     	& 255&    2   & 60\%   \\        
 \hline
\end{tabular}
  \end{center}
\end{table}


\subsection{Analyses and Specification of Hyperparameters}

We fit the proposed GP-RJ models to the four real-life survival datasets. In the absence of further information, we set the Poisson rate parameter $\alpha$ to 5, 10, 20, which reflected an \emph{a priori} expectation of 6, 11, 21 intervals on hazard time scale, respectively. For illustration, we also presented the analyses based on GP-EQ models where the time partition was prespecified with three different values of $J$= 5, 10, 20. Additional analyses were conducted with a fixed time partition constructed based on uniquely ordered event times, which we referred to as GP-UQ. For all of GP-RJ, GP-EQ, GP-UQ, we set $(\eta_0, \kappa_0, c_0) = (0.2, 0.5, 1)$ as done in Section \ref{sec:simulation}. 

Results were based on samples from the joint posterior distribution obtained from two independent MCMC chains. Each chain was run for 1 million iterations with the first half taken as burn-in. We assessed convergence of the Markov chains through the calculation of the PSRF requiring the factors to be less than 1.05 for all model parameters. The overall acceptance rates for the MH and MHG steps in the MCMC scheme ranged between 30\% and 40\%, indicating that the algorithm was relatively efficient.

\begin{table}[ht]
\centering
\caption{DIC and LPML for seven models fit to the four datasets. The smallest DIC and the largest LPML values are highlighted in bold. \label{tab:DIC}}
\begin{tabular}{l cc c cc c cc c cc}
  \hline
  &\multicolumn{2}{c}{DLC} && \multicolumn{2}{c}{GBC} && \multicolumn{2}{c}{NBC} && \multicolumn{2}{c}{ABC} \\  \cline{2-3}\cline{5-6}\cline{8-9} \cline{11-12}  
 & DIC & LPML & & DIC & LPML &  & DIC & LPML &  & DIC & LPML \\ 
  \hline
GP-EQ, $J=5$ 		& 38.5 & -19.9 &  		& 510.9 & -256.6 &  			& 187.3 & -93.3 &  			& 207.2 & -103.6 \\ 
GP-EQ, $J=10$ 	& 38.4 & -19.8 &  		& 508.0 & -255.2 &  			& 179.4 & -90.1 &  			& 206.5 & -103.3 \\ 
GP-EQ, $J=20$ 	& 39.2 & -20.3 &  		& 509.6 & -255.7 &  			& 187.5 & -92.9 &  			& 207.0 & -103.5 \\ 
GP-UQ$^{\dag}$ 	& 39.9 & -20.5 &  		& 515.9 & -259.0 &  			& 189.1 & -93.5 &  			& 207.1 & -103.6 \\ 
GP-RJ, $\alpha=5$ 	& 38.1 & -19.6 &  		& 509.7 & -255.9 &  			& 180.9 & -89.4 &  			& 205.9 & -103.0 \\ 
GP-RJ, $\alpha=10$ & 38.1 & -19.6 &  		& 506.8 & -254.4 &  			& \bf{178.0} & \bf{-88.2} &  	& \bf{205.6} & \bf{-102.8} \\ 
GP-RJ, $\alpha=20$ & \bf{37.0} & \bf{-19.1} &  	& \bf{506.6} & \bf{-254.4} &  	& 179.7 & -89.3 &  			& 206.1 & -103.0 \\ 
   \hline
   \multicolumn{12}{l}{\footnotesize$\dag$ $\bfs$ is set to uniquely ordered event times. $J=34, 270, 48, 103$ for DLC, GBC, NBC, ABC, respectively.}\\      
\end{tabular}
\end{table}

\begin{figure}[hp]
\centering
\includegraphics[width = 5in]{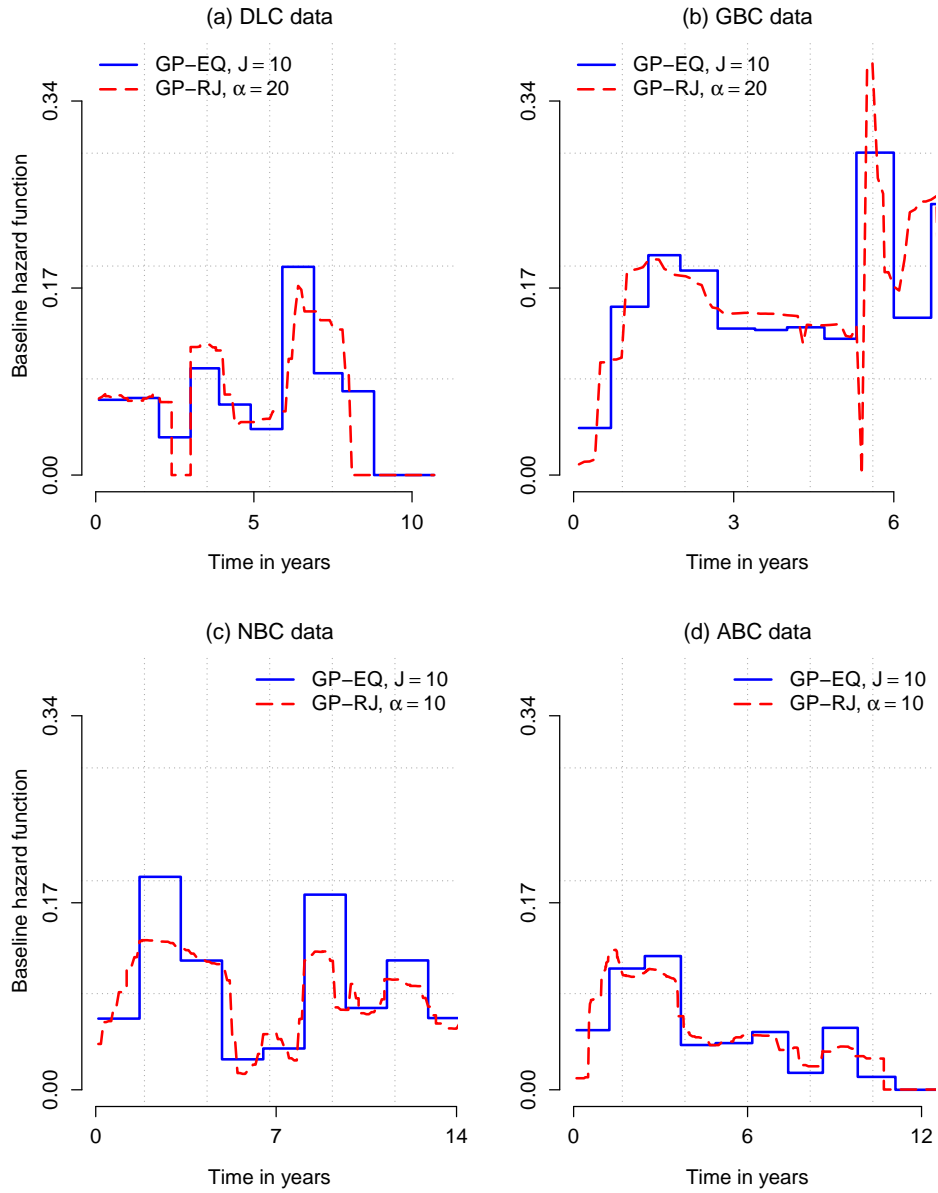}
\caption{Estimated baseline hazard function, $h_0(t)$, for the two models fit to the four real-life survival datasets outlined in Table \ref{tbl:dataSummary}.}\label{figure:dataBH}
\end{figure}

\begin{figure}[hp]
\centering
\includegraphics[width = 6in]{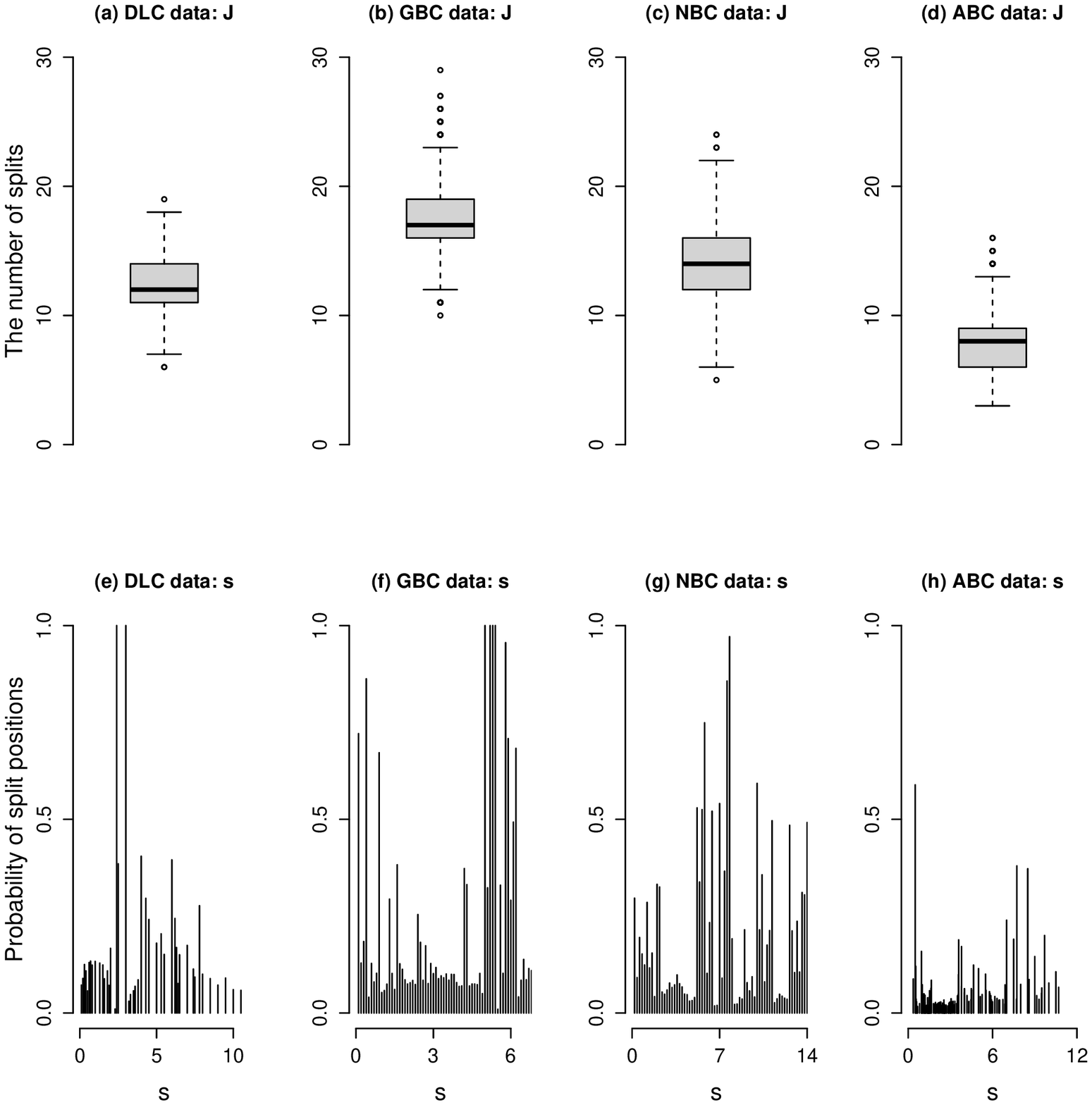}
\caption{Analyses of the four real-life survival datasets outlined in Table \ref{tbl:dataSummary}: (a)-(d) The posterior distribution of $J$; (e)-(h) The posterior probabilities associated with the positions of the split times $\bfs$.}\label{figure:dataJs}
\end{figure}

\subsection{Results}

Table \ref{tab:DIC} presented DIC and LPML for the models we compared. Focusing on the analyses where $J$ for GP-EQ and $\alpha$ for GP-RJ were set to the same value, GP-RJ model had equal or better fits to the four datasets than the corresponding GP-EQ model: e.g. differences in DIC ranged between $0.3-2.2$ for DLC data, $1.2-3.0$ for GBC data, $1.4-7.8$ for NBC data, and $0.9-1.3$ for ABC data. This implied that GP-RJ generally provided an equal or better model fit in the absence of the prior information on the time scale. 

In Figure \ref{figure:dataBH}, we presented estimates of the baseline hazard function. We provided the results from the analyses using GP-RJ and GP-EQ that had the best model fits based on DIC and LPML for each dataset. As seen in Figure \ref{figure:dataBH}, the estimated baseline hazard functions were substantially smoothed out in the GP-RJ analyses. In terms of computation speed, GP-RJ ran $1.2-1.5$ times faster than the corresponding GP-EQ in the analyses of the DLC, GBC, and ABD datasets even though GP-RJ required the extra MHG steps to update ($J$, $\bfs$). This was not surprising because the posterior medians of $J$ from GP-RJ were 12, 17, and 8 for the three datasets (see Figure \ref{figure:dataJs} (a), (b), and (d)), which were smaller than the prior Poisson rate $\alpha$'s set for the analyses. In Figure \ref{figure:dataJs} (e)-(h), we presented posterior probabilities associated with the positions of the split times $\bfs$. Note that the shape of the distribution of posterior probabilities of $\bfs$ was generally consistent with the smooth or abrupt changes in the pattern of baseline hazard (Figure \ref{figure:dataBH}).

\section{Discussion} \label{sec:discussion}

We have described a new Bayesian framework for proportional hazards models where the cumulative baseline hazard function was modeled \emph{a priori} by a gamma process. The proposed approach helps avoid the difficult task of specifying the number of the time splits and their positions by employing a random split-times model and a reversible jump MCMC scheme. 

When compared to the model with the fixed time partition (GP-EQ), the proposed GP-RJ characterized the baseline hazard function as a notably smoother mixture of piecewise constant (Figure \ref{figure:simBH} and \ref{figure:dataBH}). Although requiring the extra steps to estimate ($J$, $s$), GP-RJ generally performed well with a smaller number of $J$ (Figure \ref{figure:simBH} (e) and \ref{figure:dataJs} (a)-(d)), exhibited a smaller bias for regression coefficients (Table \ref{tab:simReg}), and yielded a better model fit (Table \ref{tab:DIC}) than the corresponding GP-EQ.

To our best knowledge, this work is the first attempt to relax the assumption of fixing the time partition for the proportional hazards model where a gamma process prior is used for cumulative baseline hazard function. The proposed framework provides researchers with valid statistical approaches to overcome a major barrier for the practical use of gamma process models in the analysis of survival data.

%

\bibliographystyle{apalike}

\bibliography{BSurvGP}
\end{document}